\documentclass[a4paper]{article}

\usepackage{INTERSPEECH2021,balance}

\usepackage{xcolor}

\title{PANACEA cough sound-based diagnosis of COVID-19\\ for the DiCOVA 2021 Challenge}
\name{Madhu R. Kamble$^1$, Jose A. Gonzalez-Lopez$^2$, Teresa Grau$^3$, Juan M. Espin$^3$, Lorenzo Cascioli$^1$, Yiqing Huang$^1$, Alejandro Gomez-Alanis$^2$, Jose Patino$^1$, Roberto Font$^3$, Antonio M. Peinado$^2$, Angel M. Gomez$^2$, Nicholas Evans$^1$, Maria A. Zuluaga$^1$, Massimiliano Todisco$^1$}

\address{$^1$EURECOM, France\\  $^2$University of Granada, Spain\\ $^3$Biometric Vox S.L., Spain}

 \email{\{name.surname\}@eurecom.fr \\ \{joseangl, agomezalanis, amp, amgg\}@ugr.es \\ 
 \{teresa.grau, jm.espin, roberto.font\}@biometricvox.com }

\begin{document}

\maketitle
\begin{abstract}

The COVID-19 pandemic has led to the saturation of public health services worldwide. In this scenario, the early diagnosis of SARS-Cov-2 infections can help to stop or slow the spread of the virus and to manage the demand upon health services. This is especially important when resources are also being stretched by heightened demand linked to other seasonal diseases, such as the flu. In this context, the organisers of the DiCOVA 2021 challenge have collected a database with the aim of diagnosing COVID-19 through the use of coughing audio samples. 
This work presents the details of the automatic system for COVID-19 detection from cough recordings presented by team PANACEA. This team consists of researchers from two European academic institutions and one company: EURECOM (France), University of Granada (Spain), and Biometric Vox S.L. (Spain). 
We developed several systems based on established signal processing and machine learning methods. Our best system employs a Teager energy operator cepstral coefficients (TECCs) based front-end and Light gradient boosting machine (LightGBM) back-end. The AUC obtained by this system on the test set is 76.31\% which corresponds to a 10\% improvement over the official baseline. 

\end{abstract}

\section{Introduction}

A year ago the COVID-19 pandemic caused a significant health crisis. COVID-19 is provoked by the infection with the severe acute respiratory syndrome virus strain SARS-CoV-2. According to the World Health Organization (WHO), the most common symptoms of COVID-19 are fever, dry cough and shortness of breath. The WHO mission report in China~\cite{WHO_report_2019} has described the symptoms of this disease from more than 55,000 laboratory-confirmed cases. These symptoms include: fever (87.9\%), dry cough (67.7\%), fatigue (38.1\%), sputum production (33.4\%), shortness of breath (18.6\%), sore throat (13.9\%), headache (13.6\%), myalgia or arthralgia (14.8\%), chills (11.4\%), nausea or vomiting (5.0\%), nasal congestion (4.8\%), diarrhea (3.7\%), hemoptysis (0.9\%), and conjunctival congestion (0.8\%). Among these symptoms, there is a significant percentage of alterations related to the respiratory system as a consequence of the infections caused by the coronavirus, which can lead to severe pneumonia~\cite{Sohrabi_2020}. 

These symptoms lead us to venture the hypothesis that it would be in principle possible to detect COVID-19 through a person's altered respiratory patterns. Thus, a recent literature review on radiological data in patients with COVID-19~\cite{Lomoro2020} concluded that these patients present abnormalities in chest radiographic images that are characteristic of this disease.  It is therefore conceivable that the distinctive alterations produced by the coronavirus in the lungs will also be reflected in the respiratory patterns of patients.

The above hypothesis is the starting point of the DiCOVA 2021 challenge~\cite{2021dicovaChallenge}, which aims at developing automatic methods for diagnosing COVID-19 through the use of sound audio samples. The challenge provides a dataset with cough recordings collected from COVID-19 positive and negative individuals for a two-class classification task. These recordings were collected via crowdsourcing from multiple countries, through a website application. The challenge features two tracks: Track-1 focuses on diagnosing COVID-19 using cough sounds, while Track-2 focused on a collection of breath, sustained vowel phonation, and number counting speech recordings.

In this paper, we describe our system for automatic COVID-19 detection presented to Track-1 of the DiCOVA 2021 challenge. Our system uses a perceptually-motivated front-end, parametrizing the cough recordings as sequences of Teager energy operator cepstral coefficients (TECCs)~\cite{kamble2019analysis}, along with a state-of-the-art gradient boosting classifier, namely a Light gradient boosting machine (LightGBM)~\cite{Ke2017LightGBM}.

The remainder of this paper is organized as follows. Section~\ref{sect:data} describes the datasets used during the development of our system, including the Track-1 DiCOVA challenge dataset. In Section~\ref{sec:system} the technical details of our system for COVID-19 detection are shown. Experimental setup and results are presented in Section~\ref{sect:exp_setup} and Section~\ref{sect:results}, respectively. Finally, the main conclusions of this work and future research lines are drawn in Section~\ref{sect:conlusions}.

\section{Data Resources}\label{sect:data}

In this section, we briefly describe the databases used in our experiments: the DiCOVA Challenge dataset and the COUGHVID corpus.

\subsection{DiCOVA dataset}

The DiCOVA Challenge~\cite{2021dicovaChallenge} features two tracks: Track-1 is focused on cough sound recordings while Track-2 considers cough, breath, sustained phonation, and continuous speech sound recordings. In both cases, the data is derived from Project Coswara~\cite{sharma2020coswara}, a crowd-sourced dataset of sound recordings from COVID-19 positive and non-COVID-19 individuals collected using a web-application.

The training/validation set for the Track-1 provided by the organizers contains 1040 audio files stored in .FLAC format at 44.1 kHz sampling rate. Each audio file corresponds to a unique subject. This set comprises a total of ~1.36h of cough audio recordings from 75 COVID-19-positive subjects and 965 non-COVID-19 subjects. Some metadata of the subjects is provided in a CSV file, like COVID-19 status (p/n), gender (m/f) and nationality (Indian/other). Validation can be performed using 5-fold cross validation using training and validation lists, for each of the 5 folds, that are also provided by organisers. The test set consists of 233 audio files with the same format as the training/validation set but with unknown COVID-19 status.
The statistics of the database is reported in Table~\ref{Tab:Database2017}.
\begin{table}[t]
	\centering
	\caption{Statistics of DiCOVA 2021 Challenge Database }
	\begin{tabular}{ *{4}{c}}
		\hline\hline
		Subset & \multicolumn{3}{c}{ \#  COVID-19}  \\ \cline{2-4} 
		&  Negative & Positive &Total\\
		\hline  
		Training   & 772 & 50 &822\\
		Validation & 193	&25 &  218\\	
	Test  & - &  -&233\\ \cline{1-4} 
		%		Total&42&3566&14466\\ \cline{1-4} 
		%		Total & 42  & 3566 & 15380 \\ 
		\hline\hline
	\end{tabular}
	\label{Tab:Database2017}
	%\vspace{-0.1 cm}
\end{table}

The Track-2 dataset provided for the challenge is composed of three kinds of sound recordings from each individual: breathing, sustained phonation (vowel-e) and speech (1-20 digit counting). The dataset contains 1199 audio files for each kind of sound (80 positives and 1119 negatives). A CSV file with metadata and lists for 5 training/validation folds, as in Track-1, are provided.

\subsection{COUGHVID corpus}
The COUGHVID~\cite{Orlandic2020TheCC, coughvidepfl} corpus is a crowdsourced and publicly-available dataset with over 20,000 cough recordings representing a wide range of subject ages, genders, geographic locations, and COVID-19 statuses. Experienced pulmonologists have labelled more than 2,000 recordings to determine which samples are likely to originate from COVID-19 patients.

\section{System Description} \label{sec:system}
Considering the relatively small size of the DiCOVA dataset and, in particular, the limited number of positive samples, we started by exploring transfer learning approaches in order to leverage pre-existing models trained on large datasets, although for a different task. In particular, we used neural networks trained for speaker recognition. These networks are usually used to compute utterance-level embeddings also known in the speaker recognition literature as x-vectors \cite{Snyder2018XVectorsRD}. We explored two different approaches to transfer learning: a) extracting utterance embeddings from the neural network and using them to train a binary classifier, and b) fine tuning the neural network for the task at hand. Although this seems to be a promising approach, it was unsuccessful on the DiCOVA dataset. For this reason, we focused on a perceptually-motivated front-end based on TECC features and different ensemble methods as back-end classifiers.

\begin{figure}
    \centering
    \includegraphics[width=.8 \linewidth]{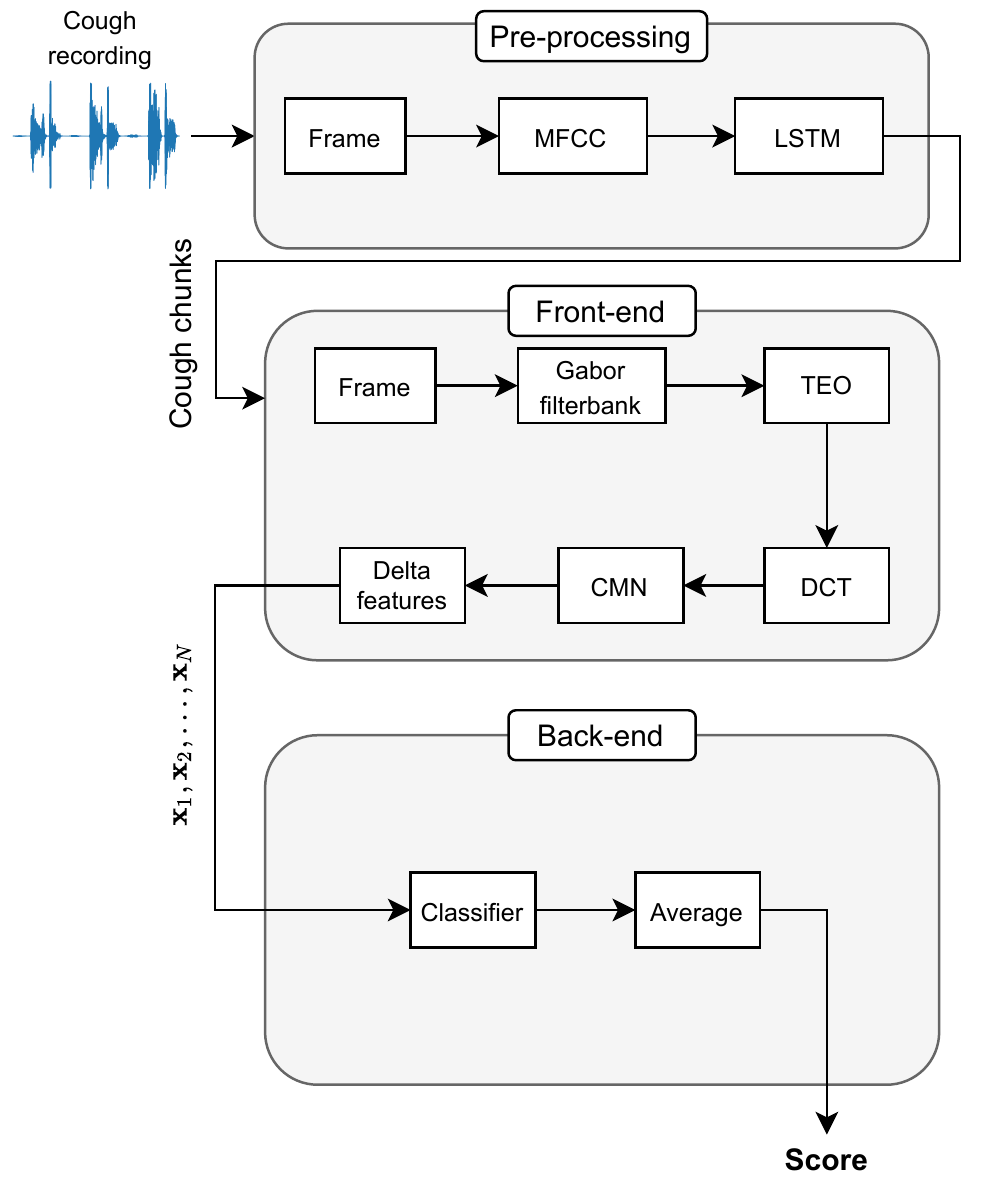}
    \caption{Block diagram of the PANACEA system for COVID-19 detection from cough recordings.}
    \label{fig:diagram}
   \end{figure}

In the following subsections, we present the technical details of our submission to the DiCOVA 2021 challenge. A block diagram of the system is shown in Figure \ref{fig:diagram}.

\subsection{Pre-processing}\label{sect:EUR_pre}

For pre-processing, a Long Short-Term Memory (LSTM)  network \cite{hochreiter1997lstm} was built to deal with a binary classification problem, with the aim of identifying whether an audio track contains cough or not. LSTMs were chosen as they are a type of neural networks which is particularly suited to the processing of sequential data like speech or video, and they proved to work as expected. The model was built on Matlab and, in practice, we trained a simple LSTM network with one hidden layer and 100 hidden units using as features the 20 MFCCs of the training audios; to train and evaluate the model, we retrieved some external cough audio files from the open-source project COUGHVID. The system was able to reach  {87\%} accuracy on the selected COUGHVID validation data. Having observed that this model was performing with acceptable precision, we exploited it to `clean' the DiCOVA dataset, deleting parts of the audios where no useful information was contained, thus extracting from each recording only the part related to cough sounds. Each audio was indeed split into small chunks of roughly one second: each chunk was then passed to the model, which outputs whether cough is present or not inside the specific part of the audio. Having done so, only chunks where cough was detected were kept and re-joined together to have a cleaned version of  original audio file.

\subsection{Front-end}

The Teager energy operator (TEO) tracks running estimate of instantaneous energy fluctuations of a narrowband speech signal \cite{maragos1991speech,maragos1992separating,maragos1993amplitude} as follows:
\begin{equation}
\Psi_d\{x_i[n]\}=x_i^2[n]-x_i[n-1]x_i[n+1]\approx a_i[n]^2\Omega_i[n]^2,
\label{Eq: ESA-TEO}
\end{equation}

where $x_i[n]$ is discrete-time, bandpass filtered signal for $i^{th}$ subband filter,  $\Psi_d\{\cdot\}$ represents TEO, $a_i[n]$  is its corresponding instantaneous 
amplitude and $\Omega_i[n]$ is instantaneous frequency. The TEO works on narrowband signal and hence, bandpass filtering is necessary to apply on the input speech signal to compute \textit{`N'} number of subband filtered signals. Here, the input speech signal is first passed through a Gabor filterbank to obtain \textit{`N'} subband filtered signals \cite{kamble2019analysis, kamble2018effectiveness}. We used Mel-spaced Gabor filterbank to have compressed bandwidth in the lower frequency region and wide bandwidth in the higher frequency regions. The narrowband filtered signals are obtained at center frequency, which are Mel-spaced between $f_{min}$=10 Hz, and  $f_{max}$=8000 Hz. These subband filtered signals are given to the TEO block to estimate the Teager energy profile of each subband filtered signals. These Teager energy profiles are further passed to the frame-blocking along with averaging of the speech segment using a window length of 25 ms and shift of 10 ms followed by logarithm operation. To obtain a low-dimensional representation that has compact energy, a Discrete Cosine Transform (DCT) is applied along with  Cepstral Mean Normalization (CMN) (also known as Cepstral Mean Subtraction (CMS)) to reduce the channel mismatch/distortion conditions \cite{molau2003feature}. Finally, the retained few DCT coefficients, i.e., Teager Energy Cepstral Coefficients (TECC) are appended along with their $\Delta$ and $\Delta\Delta$ coefficients \cite{KAMBLE2021101140}. 

\begin{figure}[!t]
	\centering
	\includegraphics[trim={1.0cm 0.5cm 0.5cm 0.5cm},clip,width=1.1\linewidth]{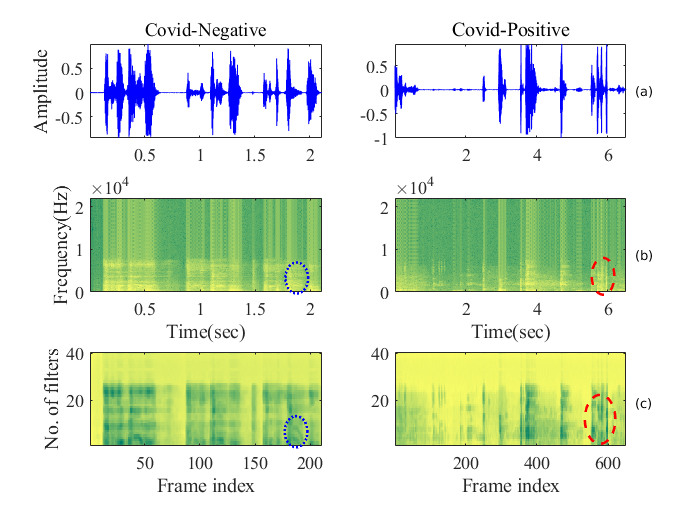}\\
	\caption{Comparison of spectral energy densities of traditional STFT spectrogram and Teager energy-based spectral features for COVID-19 negative and positive audio signals.}  
	\label{fig:spec_density}
\end{figure}

The spectral energy density obtained from 40 Mel-scaled Gabor filterbank for COVID-19 positive and negative signals is shown in Figure \ref{fig:spec_density}. We compared the spectral energy densities with the traditional short-time Fourier transform (STFT) spectrogram. The Teager energy-based spectral features preserves the formant frequencies compared to the traditional spectrogram (highlighted by blue and red dotted circles). The formant frequencies also provide valuable information related to the role of the vocal tract in the generation of an acoustic signal. It is observed from Figure \ref{fig:spec_density}, the harmonic structure for the COVID-19 negative signal shows 4 different formant frequency bands that is not present for the COVID-19 positive signal. 

\begin{figure}[!t]
	\centering
	\includegraphics[trim={4cm 7.5cm 5cm 8cm},clip,width=0.7 \linewidth]{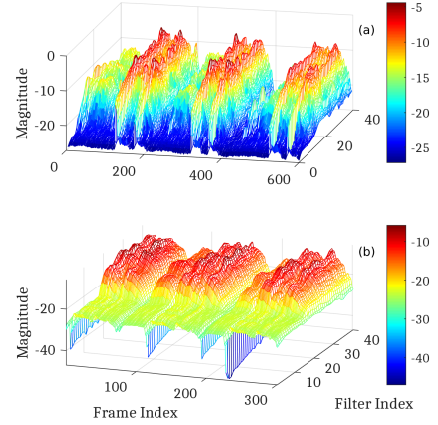}\\
	\caption{Waterfall plots for (a) COVID-19 positive and (b) COVID-19 negative audio signal.}  
	\label{fig:waterfall}
\end{figure}

Figure \ref{fig:waterfall} shows two waterfall plots for (a) COVID-19 positive  and (b) COVID-19 negative audio signals
computed from the TEO-based spectral features. This 3-dimensional pictorial representation shows the spectral spread and its magnitude range along the frequency values. It can be seen that the COVID-19 positive audio signal has higher spectral energy (indicating more red color spectral spread). This comparative analysis indicates that for the detection of COVID-19 the higher formants and frequency values are more useful.

\subsection{Back-end} \label{sect:back-end}
State-of-the-art ensemble methods were used for the task of predicting the COVID-19 status of the speaker from the TECC features extracted from the cough recordings. 

In particular, during training, a light gradient boosting machine (LightGBM) classifier \cite{Ke2017LightGBM}, which is a gradient boosting algorithm employing tree-based classifiers for classification, was trained to predict the COVID-19 status for each of the acoustic feature vectors in the training dataset. During evaluation, the COVID-19 score for each speaker was computed by averaging the scores computed by the classifier for each of the feature vectors extracted from the cough recording for that particular speaker.

\section{Experimental Setup}\label{sect:exp_setup}
\textbf{Dataset}: We employed the DiCOVA 2021 Challenge database, as discussed in Section \ref{sect:data}. 

\noindent \textbf{Baseline system}: The Challenge provides a baseline system for Task-1 based on 39-dimensional Mel-frequency cepstral coefficients (MFCCs) with $\Delta$ and $\Delta\Delta$ coefficients. Three back-end classifiers are used, namely, Logistic regression (LR), Multi-layer perceptron (MLP) and Random Forest (RF). As in our back-end model, frame-level probability scores are computed using the trained model. Finally, all the frame scores are averaged to obtain a single COVID-19 probability score for the cough recording.

\noindent \textbf{Evaluation metrics}: Classification performance evaluation is measured using traditional detection metrics, namely, true positive rate (TPR) and false positive rate (FPR) over a range of decision thresholds. From these metrics, the probability scores for each audio file are used to compute the receiver operating characteristic (ROC) curve, and the area under the curve (AUC) metric to quantify the model performance.

\noindent \textbf{Implementation details}: TECC feature vectors were extracted using 40 Mel-spaced Gabor filterbank with $f_{min}$=10 Hz, and $f_{max}$=8000 Hz. For each subband filtered signals, we obtain \textit{40}-D static features augmented with their $\Delta$ and $\Delta\Delta$ coefficients resulting in \textit{120}-D feature vector. Cepstral Mean Normalization (CMN) was applied to enhance robustness against channel mismatches. The LightGBM model was trained with 100 trees in the forest. Furthermore, Bayesian optimization was used, in particular the tree of Parzen estimators (TPE) algorithm described in \cite{Bergstra2013hyperopt}, to optimize the hyper-parameters of the LightGBM model. This procedure was applied using a 4-fold stratified cross-validation scheme, leading to improved results.

\section{Experimental Results}\label{sect:results}

\subsection{Preliminary results}

We evaluated the effect of using different acoustic features on classification performance using the RF classifier. The performance metrics for these preliminary experiments are shown in Table \ref{tab:EUR_results}. We compared our TECC features with the MFCC features in the DiCOVA baseline system (BL). On validation set, TECCs did not gave better performance, resulting an AUC of 67.28\%, however, on test set the AUC was 72.53\% whereas the AUC of the MFCC baseline was 69.85\%. Furthermore, we performed experiments using the pre-processed data as discussed in Section \ref{sect:EUR_pre}. As shown in Table~\ref{tab:EUR_results}, although the validation results obtained with the pre-processed data were better, this procedure, unfortunately, resulted in a significantly lower performance on the test set. 
To improve the AUC further, we used a score-level fusion of the MFCC- and TECC- based systems, which increased the performance up to 73.75\% AUC on test set.
\begin{table}[t]
\caption{Preliminary results on DiCOVA validation and test sets.}
\label{tab:EUR_results}
\centering
\resizebox{\columnwidth}{!}{%
	\begin{tabular}{ *{6}{|c|} }
\hline
 \begin{tabular}[c]{@{}c@{}}System \\ Description\end{tabular} & \begin{tabular}[c]{@{}c@{}}Pre- \\ Processing\end{tabular} & \begin{tabular}[c]{@{}c@{}}Avg. Val.\\  AUC\end{tabular} & \begin{tabular}[c]{@{}c@{}}Test \\ AUC\end{tabular} & Test Sens. & Test Spec. \\ \hline\hline
BL: MFCC     & No     &     68.54         & 69.85    & 80.49      & 53.65      \\ \hline
S1:TECC  & No        & 67.28         & 72.53    & 80.49      & 56.77      \\ \hline
TECC  & Yes        &   69.00     &  62.36   & 80.49      &   43.75    \\ \hline
   BL+S1 score fusion   & No          & 67.53         & \textbf{73.75}    & 80.49      & 54.69      \\ \hline
\end{tabular}}
\end{table}

\begin{figure}[!h]
	\centering
	\includegraphics[width=\linewidth]{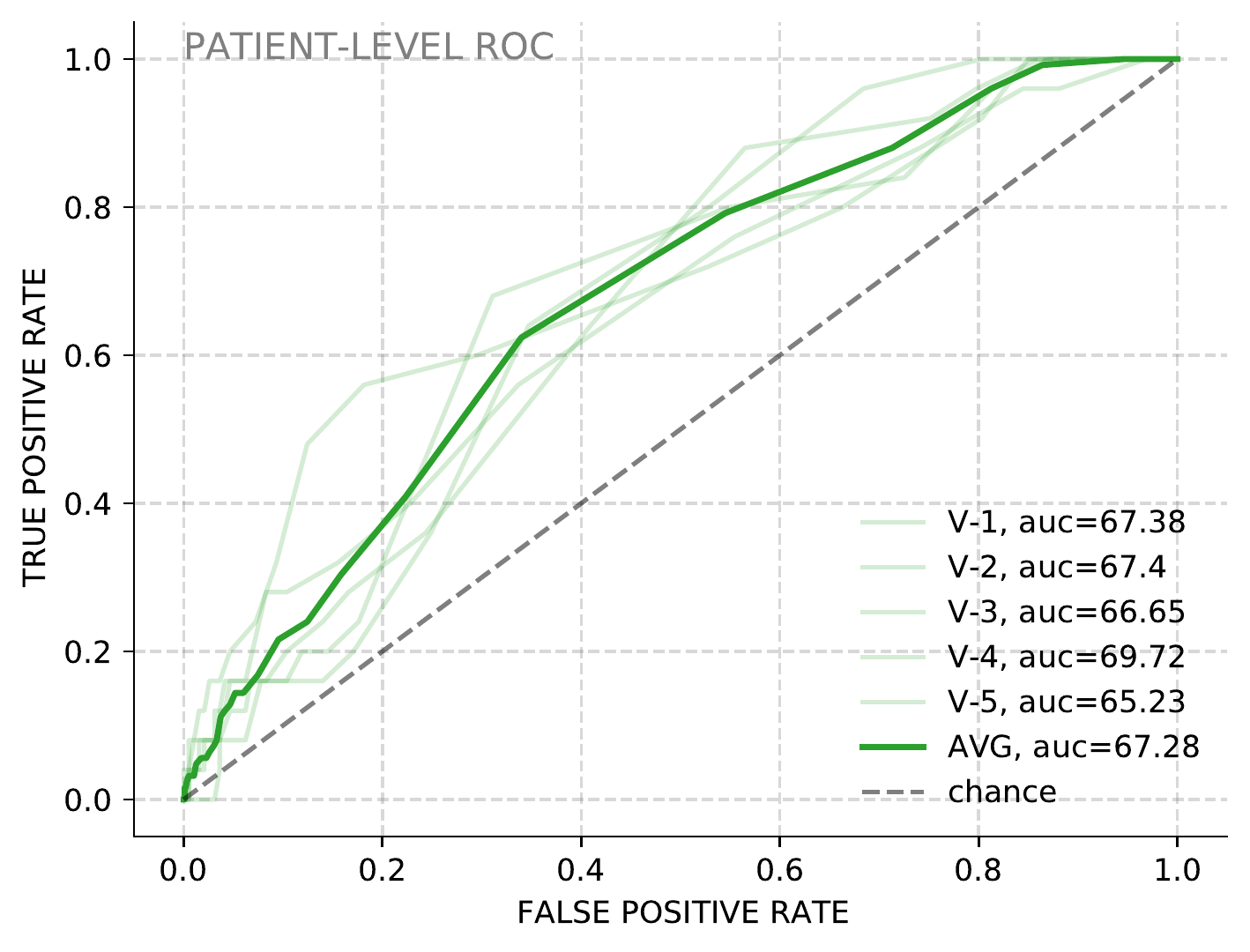}\\
% 	\vspace{-0.3cm}
	\caption{Average ROC computed for TECC features on the training/validation dataset using the RF classifier (shade indicates standard error in ROC over the 5 validation folds).}  
	\label{fig:roc}
%  	\vspace{-0.3cm}
\end{figure}

Figure \ref{fig:roc} shows the ROC curve obtained for TECC-based system with RF classifier with no pre-processing. The ROCs corresponds to the average value of all the five validation folds defined in the DiCOVA challenge. The individual AUC for each fold of validation (V-1 to V-5) are in the range of 65-70\% that shows that TECC features gave almost equal performance across all the folds. On the other hand, the MFCC features shows huge variation in AUC for 5 folds which is not a good case for the real-time data.

\subsection{Final results}
Our best performing system submitted to the DiCOVA 2021 challenge uses a LightGBM back-end, as explained in Section~\ref{sect:back-end}. Table \ref{tab:ugr_results} shows the results obtained when training this classifier with different feature sets. As can be seen, while MFCCs outperforms TECCs features on the validation set, the latter feature set obtains significantly better results on the test set. In particular, our final system submitted to the DiCOVA 2021 obtained an AUC of 76.31 \% on the test set, which places our team on the $15^{th}$ position of the official ranking.

\begin{table}[t]
\caption{Final results on DiCOVA validation and test sets with the LightGBM classifier.}
% 	\vspace{-0.2cm}
\label{tab:ugr_results}
\centering
\resizebox{\columnwidth}{!}{%
	\begin{tabular}{ *{5}{|c|} }
\hline
Feature set & \begin{tabular}[c]{@{}c@{}}Avg. Val.\\  AUC\end{tabular} & \begin{tabular}[c]{@{}c@{}}Test \\ AUC\end{tabular} & Test Sens. & Test Spec. \\ \hline \hline
39 MFCCs + $\Delta$ + $\Delta\Delta$ + CMN  &74.59         &62.96    & 80.49      & 45.83       \\ \hline    
40 TECCs + $\Delta$ + $\Delta\Delta$ + CMN   & 69.80        & \textbf{76.31}&    80.49      &  53.65     \\ \hline
\end{tabular}}
%  	\vspace{-0.5cm}
\end{table}

\section{Conclusion}\label{sect:conlusions}

We have presented the systems developed by team PANACEA for the DiCOVA 2021 challenge. These systems explore different features, back-end classifiers and transfer learning methods. Our best system, using TECC features and a LigthGBM classifier as back-end, obtains an AUC of 76.31\% on the test set, which represents a significant improvement over the baseline. Although there is still a lot of room for improvement, we do believe that these are promising results that support the idea that there are alterations in the respiratory patterns, caused by COVID-19 infection, that can be detected from cough or speech samples. Automatic analysis of such samples, that can be provided by the patient from their own home safely and noninvasively, could indeed be a powerful tool for screening and detection of COVID-19.

The small number of positive samples and the crowdsourced nature of the data used for the challenge should raise, however, some concerns about the ability of these findings and classifiers to generalize to new, different data. Since good generalization is an essential ability for these systems to be useful for any real-world scenario, our future work will focus on assessing generalization by working with both larger, more diverse datasets, for system training, and more curated data, with labels linked to gold-standard PCR result, for system evaluation.

\section{Acknowledgements}
Jose A. Gonzalez-Lopez holds a Juan de la Cierva-Incorporation Fellowship from the Spanish Ministry of Science, Innovation and Universities (IJCI-2017-32926). The work is also supported by the RESPECT project funded by the French Agence Nationale de la Recherche (ANR) and the German Research Foundation Deutsche Forschungsgemeinschaft (DFG), and the Spanish Ministry of Science, Innovation and Universities Project No. PID2019-104206GB-I00/SRA(State Research Agency)/10.13039/501100011033.

\newpage
\balance
\bibliographystyle{IEEEtran}

\bibliography{template}

\end{document}